\documentclass[useAMS,usenatbib]{mn2e}

\usepackage{psfig}
\usepackage{amsmath}
\usepackage{amssymb}
\usepackage{upgreek}
\usepackage{longtable}
\usepackage[mathscr]{eucal}
\usepackage{graphicx}
\usepackage{float}
\usepackage{subfig}
\usepackage[export]{adjustbox}
\usepackage[usenames, dvipsnames]{color}
\usepackage{titlesec}
\usepackage{csvsimple}
\titlespacing*{\section}{0pt}{1.1\baselineskip}{\baselineskip}
\titlespacing*{\subsection}{0pt}{1.1\baselineskip}{\baselineskip}

\title
[The dynamics of the $\gamma$ Vel cluster and nearby Vela OB2 association]
{The dynamics of the $\gamma$ Vel cluster and nearby Vela OB2 association}
\author
[Armstrong et al.]
{Joseph J. Armstrong$^{1}$, Nicholas J. Wright$^{1}$, R. D. Jeffries$^{1}$ and R. J. Jackson$^{1}$\\
$^{1}$Astrophysics Group, Keele University, Keele, ST5 5BG, UK
}

\begin{document}
\maketitle

\begin{abstract}
The kinematics of low-mass stars in nearby OB associations can provide clues about their origins and evolution. Combining the precise positions, proper motions and parallaxes given in the second Gaia Data Release with radial velocity measurements obtained with the Hermes spectrograph at the Anglo-Australian Telescope, we have an opportunity to study in detail the kinematics of low-mass stars belonging to the nearby $\gamma$ Vel cluster and the Vela OB2 association it is projected against. The presence of lithium is used to confirm the youth of our targets. We separate our sample into the cluster and association populations based on the membership probabilities of \citet{jeffries14}, their parallaxes, and kinematics. We find strong evidence for expansion in the OB association population with at least 4$\sigma$ significance along all three axes, though the expansion is notably anisotropic. We discuss these results in the context of cluster and association dispersal theories.
\end{abstract}


\section{Introduction}
\indent OB associations are sparse groups of kinematically associated but gravitationally unbound stars. Their brightest members have been studied for many years \citep{ambartsumian47,blaauw64}, but it is only recently that low-mass stars belonging to these associations have begun to be identified over large areas \citep{preibisch02,briceno07,armstrong18,cantatgaudin19b}. \\
\indent One hypothesis explaining the origin of OB associations is that they are the unbound remnants of previously bound young clusters \citep{tutukov78,hills80,kroupa01b}. After a few Myrs, newly formed O and B type stars sweep away the molecular cloud in which the cluster is embedded via feedback from photoionizing radiation and stellar winds, and the loss of this binding mass causes the cluster to become unbound and disperse. However, more recent work using astrometric data from Gaia \citep{gaia16a} has begun to challenge this hypothesis. \citet{wright18} analysed the dynamics of the Scorpius-Centaurus OB2 association and found that it did not display the radial expansion pattern expected if the association had been formed as a more compact cluster (or clusters). Rather, they concluded that the association was most likely formed in multiple highly-substructured subgroups, a view supported by the age distribution found by \citet{pecaut16}. \citet{ward18} examined kinematic parameters of 18 nearby associations and concluded that none of these associations showed signs of evolving from clusters. This new evidence implies that star formation can take place over regions of various densities and that regions of high local density form gravitationally bound clusters while low density regions form unbound associations \citep{kruij12}. \citet{kuhn19} investigate the kinematics of 28 clusters and associations by looking at young stellar objects (YSOs) in Gaia DR2 and find that at least 75\% shows signatures of expansion with a median velocity of $\sim$0.5 kms$^{-1}$. Their results indicate that some young clusters can contain significant substructure and do still exhibit the potential to expand to the scales of OB associations.  \\ 
\indent The $\gamma$ Velorum cluster is a dense group of young stars \citep[10 - 20 Myr;][]{pozzo00,jeffries09,jeffries17} located in the Vela OB2 complex \citep[at a distance of $\sim410$pc;][]{dezeeuw99} which includes many lithium-rich, pre-main sequence stars. The study by \citet{jeffries14} using radial velocities (RVs) from the Gaia-ESO survey (GES) \citep{gilmore12,randich13} identified two kinematically distinct populations within their sample of stars towards the cluster, population A, a compact component with a narrow 1D RV dispersion $\sigma_A = 0.34 \pm 0.16$ kms$^{-1}$ and a potentially more widespread population B with a broader RV dispersion $\sigma_B = 1.60 \pm 0.37$ kms$^{-1}$, whose mean RVs are offset by $2.15 \pm 0.48$ kms$^{-1}$. \citet{sacco15} established that the $\sim35$ Myr cluster NGC 2547, located $\sim2$ degrees south of $\gamma$ Vel, also exhibits two kinematically distinct populations, and that populations NGC 2547 B and $\gamma$ Vel B have similar RVs and lithium abundances.\\
\indent Since the release of Gaia data, the Vela OB2 region has been the subject of a number of studies focussed on identifying and studying the young population across the association \citep{damiani17,armstrong18,franciosini18,beccari18,cantatgaudin19a,cantatgaudin19b}. \citet{franciosini18} focused on the $\gamma$ Vel cluster and found that populations A and B are separate in parallax as well as RV ($\varpi_{A} = 2.85 \pm 0.008$ mas, $\varpi_{B} = 2.608 \pm 0.017$ mas), making $\gamma$ Vel A $\sim38$ pc closer, and also found an inverse-correlation between parallax and RV in $\gamma$ Vel B, which suggests that this group is expanding. \citet{beccari18} made use of the Density Based Spatial Clustering of Applications with Noise \citep[DBSCAN,][]{ester96} clustering algorithm to identify six groups of stars within a $\sim55$ deg$^2$ area, four of which are distinct in position and proper motion space including both $\gamma$ Vel and NGC 2547. These clusters correlate with the extended PMS population of Vela OB2 identified by \citet{armstrong18} using Gaia DR1 and 2MASS photometry. \citet{cantatgaudin19a} combine Gaia DR2 astrometry and GES data and select stars using photometry and the unsupervised photometric membership assignment in stellar clusters \citep[UPMASK,][]{krone-martins14} scheme to group stars based on their proper motions and parallax. Among stars of Vela OB2 identified, they find the distribution is fragmented into 11 components arranged in a ring-like structure around the IRAS Vela Shell. They consider the possibility that the expansion of the shell triggered star fomation in Vela OB2. \citet{cantatgaudin19b} then use UPMASK to group stars in a large field across the Vela-Puppis region ($> 600$ degrees). They identify 7 distinct populations of various ages (ranging from $\sim8 - 50$ Myr), all of which show signs of expansion in the Galactic X and Z dimensions and substructure in positions and kinematics. \\
\indent In order to better understand the 3D dynamical state and evolution of the wider Vela OB2 association we have undertaken a spectroscopic survey of the region. In this study we present the first results from this survey for a 1 degree radius field covering the Gamma Vel cluster and the existing GES field.

\section{Data}

\begin{figure}
\includegraphics[width=\columnwidth]{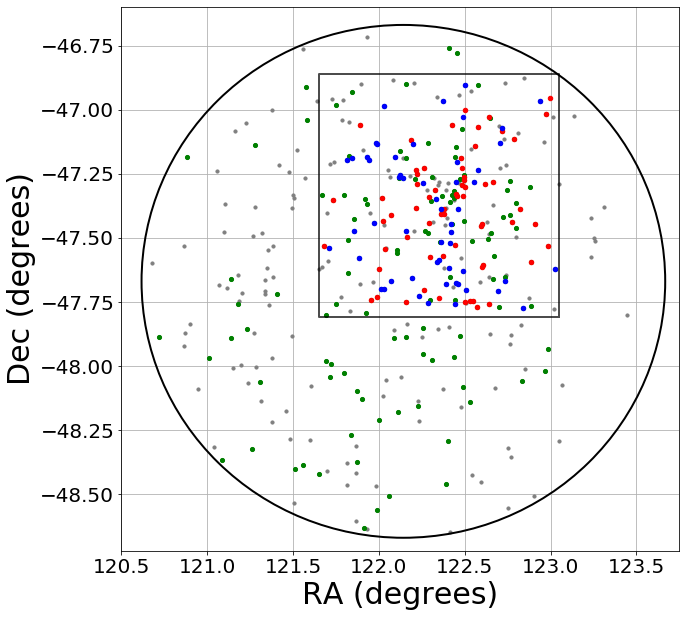}
\setlength{\belowcaptionskip}{-10pt}
\setlength{\textfloatsep}{0pt}
\caption{Positions of combined GES (square area) and AAT (circular area) sample sources. Sources with EW(Li) $>150 \,$m\AA\ (green) along with population A (red) and B (blue) members with membership probabilities $> 0.75$ from \protect\citet{jeffries14}.}
\label{CSP}
\end{figure}

\begin{figure*}
\includegraphics[width=500pt]{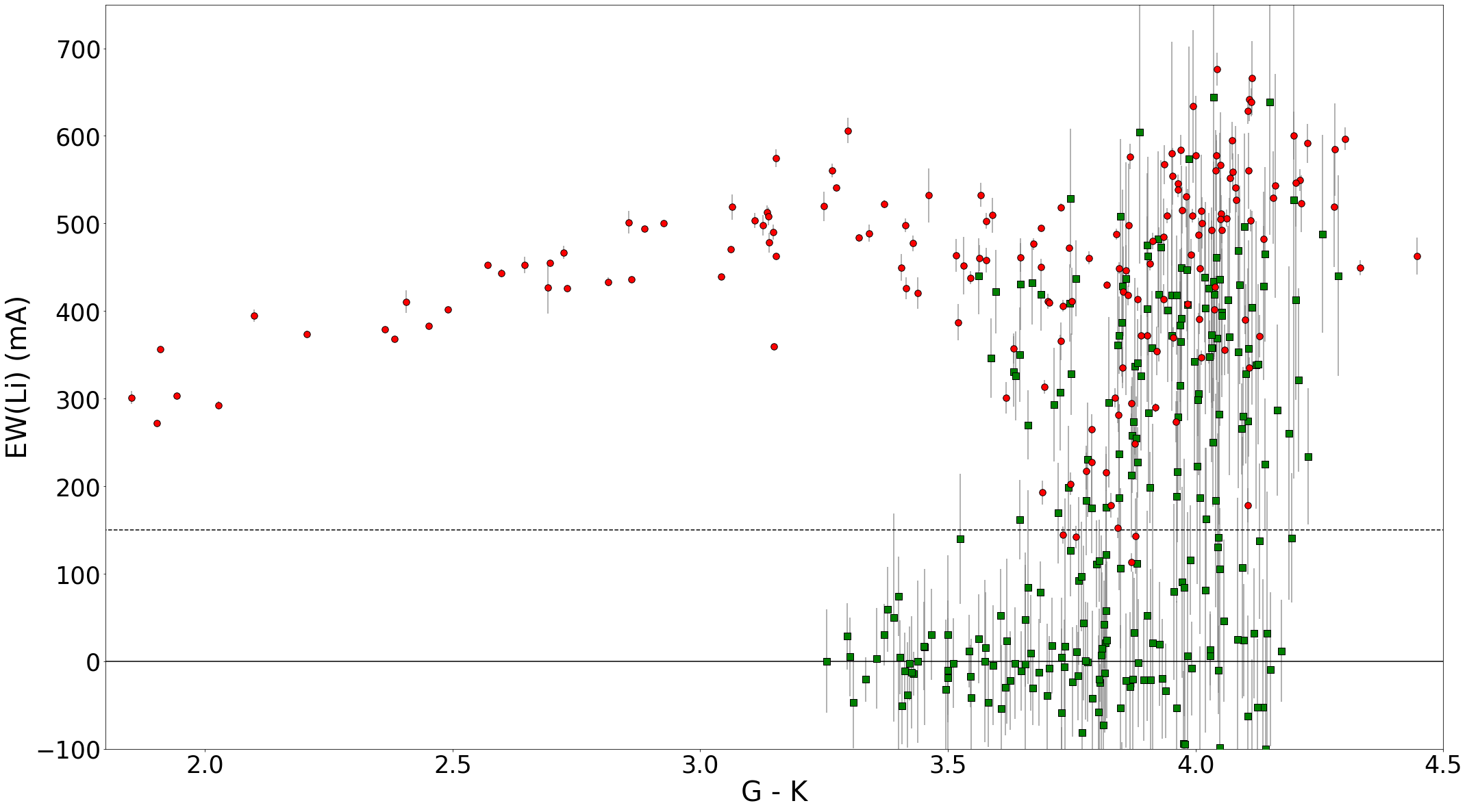}
\setlength{\belowcaptionskip}{-10pt}
\setlength{\textfloatsep}{0pt}
\caption{G - K colour versus EW(Li) for 248 AAT targets (green) with 'AAA' quality infrared photometry and EW(Li) measurements and 170 GES members (red) from \protect\citet{jeffries14} with 'AAA' quality infrared photometry. The EW(Li) dividing line at $150\,$m\AA\ is shown (see Section 2.4). }
\label{G-KvEW(Li)}
\end{figure*}

\begin{figure}
\includegraphics[width=\columnwidth]{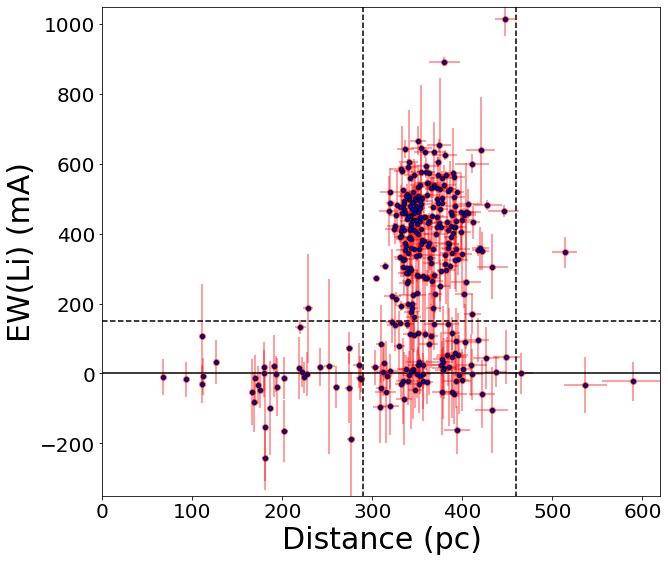}
\setlength{\belowcaptionskip}{-10pt}
\setlength{\textfloatsep}{0pt}
\caption{Distance estimates from \protect\citet{bailerjones18} versus measured EW(Li) for 327 sources with distance $<$ 600 pc from the combined sample. The clear group between 290 - 460 pc with significant EW(Li) ($>$150 m\protect\AA) are likely young stars.}
\label{DvEW(Li)}
\end{figure}

\subsection{Target selection}
We selected all Gaia DR1 sources within a 1 degree radius of (l, b) = (263$^{\circ}$, -8$^{\circ}$), which overlaps the GES $\gamma$ Vel field (see Fig. 1), within the Gaia-magnitude (G) range $14.5 - 17.5$ mag. These were then matched by position with the 2MASS catalogue \citep{skrutskie06} with a matching radius of 0.5 arcseconds. Sources with J, H, K photometry uncertainties $> 0.05$ mag or with possible contamination (as indicated by the "Cflg" flag) were excluded. We then filter these sources through a G-K vs G colour-magnitude selection and a H-K vs J-H colour-colour selection using the method described in \citet{armstrong18}, producing a sample of 360 likely PMS stars as targets for observation. 

\subsection{Observations and data reduction}
Observations were made with the 2-degree field  \citep[2dF,][]{lewis02} fibre positioner and the High Efficiency and Resolution Multi-Element Spectrograph \citep[HERMES,][]{hermes} at the Anglo-Australian Telescope (AAT) on the 6th February 2018. HERMES observes in four optical bands and the red band within which the Li~{\sc i}~6708\AA\ absorption line is found covers the wavelength range 647.8 - 673.7 nm. This is divided among 4096 pixels (0.0063 nm per pixel) on the CCD and yields a typical SNR of 100 per resolution element in 60min for a source of magnitude V = 14 \citep{hermes}. 3 x 2400s exposures were completed for the field covering the $\gamma$ Vel cluster. Calibration frames, including wavelength calibration frames, dark frames and multi-fibre flat fields were taken with the exposures to be subtracted from the target spectra during reduction \citep{lewis02}. Also, 25 fibres per field were allocated to regions of empty sky in the field to measure the 'sky spectrum', which is then subtracted from the target spectra. The spectroscopic data were reduced using the software 2dFdr \citep{2dfdr}. RVs were measured from the reduced spectra by cross-correlating the median spectra of individual targets with the spectra of standard stars and then fitting a Gaussian function to characterize the peak in the cross-correlation function, following the procedure of \citet{jackson10,jackson18}. We match our  AAT targets by position with the Vista Hemishpere Survey catalogue \citep[VHS, ][]{VHS} and combine the VHS and 2MASS K-band measures by taking the 2MASS value for K $< 12$, taking the mean of the two measures for $12 <$ K $< 13$, and then using VHS for K $> 13$. We then use Gaia DR2 G magnitude to calculate the G - K colour for these sources and perform SED fitting using the method outlined in \citet{wright19} to estimate effective temperatures ($T_{\rm eff}$). \\
\indent In order to determine EWs for the Li 6708\AA\ feature, we used a spectral subtraction technique that required template spectra of similar effective temperature ($T_{\rm eff}$) and gravity ($\log g$) to the targets (but without lithium) in order to isolate the contribution of Li. Templates were synthesised for $\log g =4.5$ at 100 K steps with a minimum of 4000 K, using the MOOG spectral synthesis code \citep{Sneden2012a}, with the \citep{Kurucz1992a} solar-metallicity model atmospheres. Equivalent widths of the Li~{\sc i}~6708\AA\ absorption line were measured by integrating under the relevant profile of the spectra after subtraction of the template. The extraction profile accounted for both the instrumental resolution, rotational broadening and offset in RV. The linelists and atmosphere models do not include the strong molecular contributions that become important at low temperatures. For that reason, the lowest $T_{\rm eff}$ used for the templates was 4000 K, which leads to a systematic (but consistent) zeropoint error in EW(Li) for stars cooler than this. However, this offset is appears to be small (see Fig. 2), and these EWs are accurate enough to enable the selection of Li-rich objects (see Section 2.4).\\
\indent We obtained spectra for all 360 targets in this field, extracting RVs and EW(Li)s for 248 (68.9\%) of these with spectroscopy of sufficient quality (SNR $>$ 5). Of these, the median uncertainty in RV is 1.88 kms$^{-1}$ and in EW(Li) is 80.26 $\,$m\AA. The EW(Li) for these 248 targets are shown in Fig.2, compared with Li-rich members of $\gamma$ Vel defined in \citet{jeffries14}.

\subsection{Compiling the sample}
We take the GES sample of the $\gamma$ Vel cluster \protect\citep{jeffries14} of 208 sources in a $0.9$ square degree area and concatenate this with our AAT sample of 248 sources. 52 sources have repeat observations in both GES and AAT samples and have measurements of RV and EW(Li) for both, so we calculated mean values weighted by the square of the inverse measurement uncertainty. We removed 8 sources where RV measurements indicated these were binary systems and measured a median offset of 1.21 kms$^{-1}$ between RV measurements for the remaining 44 sources. Since measurements from the GES sample are of higher quality than our AAT measurements, we add this median RV offset to all AAT RVs to bring the samples into agreement. At this stage we have a sample of 422 unique sources with spectroscopic RVs and EW(Li). \\
\indent On 27th April 2018 the second Gaia data release (DR2) became available, containing proper motion and parallax data for $\sim97\%$ of our sample. 12 sources lack DR2 5-parameter astrometry so we discard these. Sources were matched to the Gaia DR2 catalogue and then filtered on the suggested quality criteria to avoid using sources with spurious astrometric solutions \citep[eqs. 1,2 and 3 from][]{arenou18}. We also calculate renormalised unit weight error (RUWE) values for these sources (using Gaia DR2 RUWE data, see technical note GAIA-C3-TN-LU-LL-124-01) and discard those with RUWE $>$ 1.4 as advised by \citet{lindegren18}. Removing these leaves 339 unique sources with spectroscopic RVs, EW(Li) and 5-parameter astrometry.

\begin{figure}
\includegraphics[width=\columnwidth]{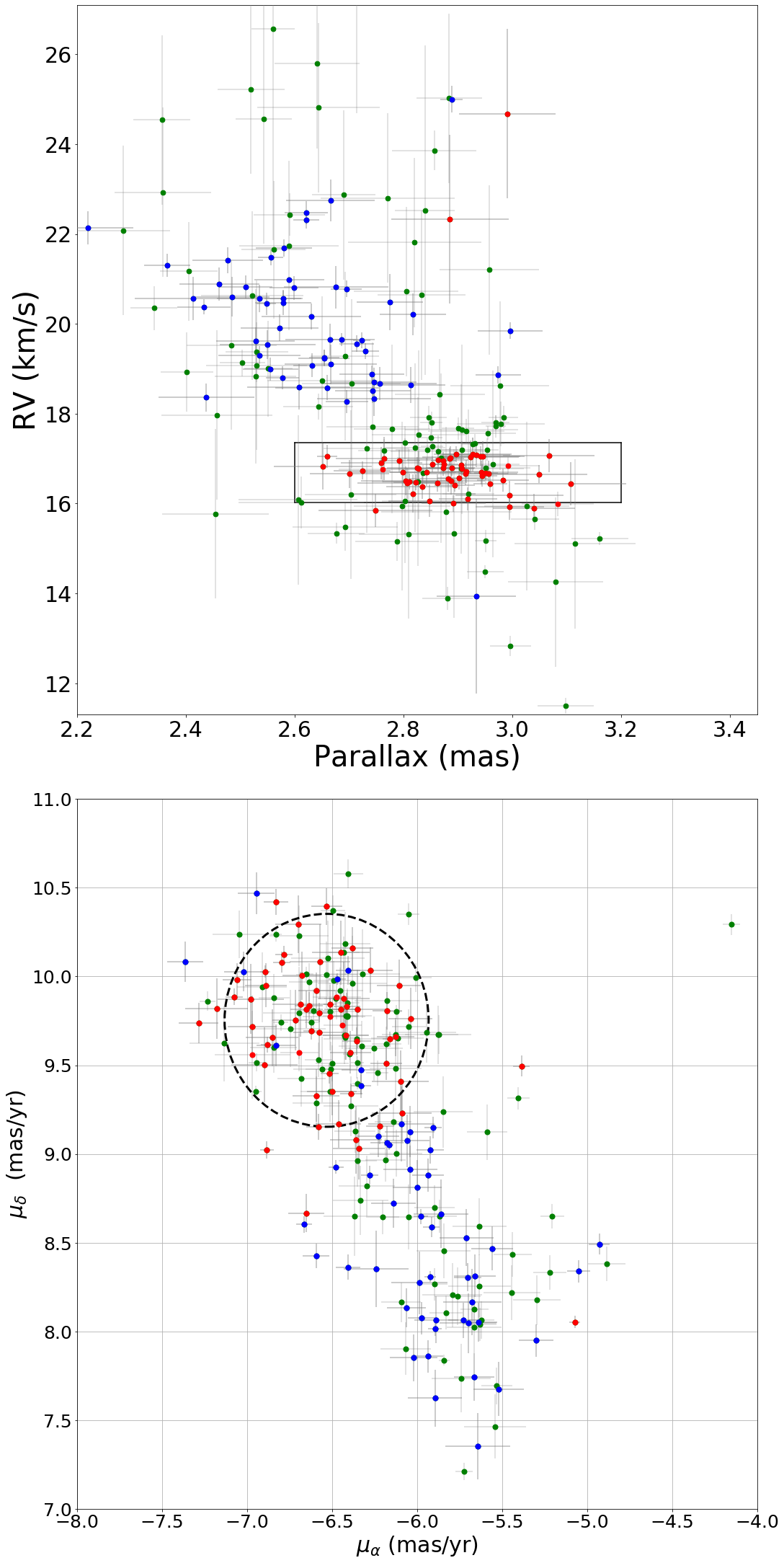}
\setlength{\belowcaptionskip}{-10pt}
\setlength{\textfloatsep}{0pt}
\caption{$Top:$ Parallax vs RV of sources with EW(Li) $>150 \,$m\AA\ (green) from our spectroscopic survey along with population A (red) and B (blue) from \protect\citet{jeffries14}, showing the divisions we use to allocate sources to each population. 15 sources are beyond the RV range shown. $Bottom:$ Gaia DR2 proper motions of the same sources.}
\label{CSplxrv}
\end{figure}

\subsection{Selection of young stars}
Distance estimates were taken from \citet{bailerjones18} for all 341 sources with clean and complete spectroscopy and astrometry and these are shown in Figure 3 for the 327 sources with distance $<$ 600 pc, plotted against their EW(Li). There is a clear group apparent at 300 - 400 pc with significantly higher EW(Li) measurements than the rest of the sample that suggests these are young stars at the distance of the $\gamma$ Vel cluster. The distinction between the young stars and the contaminating field stars becomes unclear for EW(Li) $<$ 150 $\,$m\AA. In \citet{jeffries14}, the criterion for GES sources to be considered young stars was EW(Li) $> 100 \,$m\AA, but since our AAT measurements have a lower precision than the GES data we set the threshold at $150\,$m\AA\ (see figures 2 and 3). There could be a few highly Li-depleted objects that are filtered out from the sample at this stage, though they are likely $< 10\%$ of $\gamma$ Vel cluster members \citep[][; Jackson et al. 2020 in preparation]{prisinzano16}.
\\

\section{Analysis}

\begin{figure}
\begin{center}
    \includegraphics[width=\columnwidth]{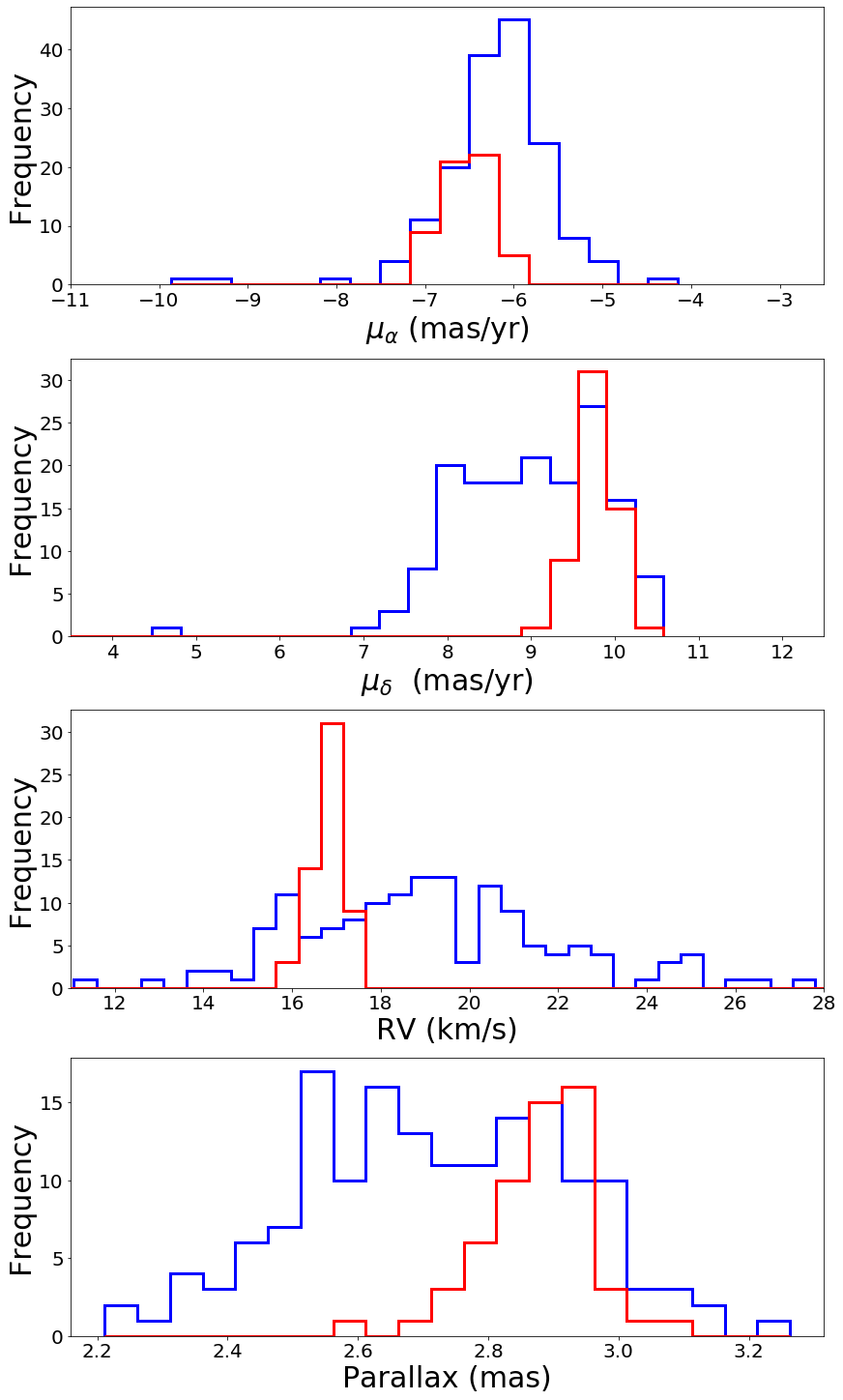}
    \setlength{\belowcaptionskip}{-10pt}
    \setlength{\textfloatsep}{0pt}
    \caption{Histograms of the velocity and parallax data for sources from our sample with EW(Li) $>150 \,$m\AA\ without 3$\sigma$ outliers in any dimension which belong to populations A (red) and B (blue). }%
    \label{SelectionBoxes}%
\end{center}
\end{figure}

\subsection{Member selection from the GES sample}
In \citet{jeffries14} the two populations were identified in their Li-rich sample using a two-component Gaussian fit to the RV distribution.
For each source, membership probablitites ($P_A, P_B$) were calculated from likelihood functions of the model RV distribution for the two populations. Sources were considered reasonably secure members of either population if either $P_A$ or $P_B$ $> 0.75$ (where $P_A + P_B =1$ , i.e. the Li-rich sample was assumed uncontaminated).
\\
\subsection{Member selection from the combined sample}
We use the \citet{jeffries14} membership information as a guide to help identify the differences in position and velocity between the two populations, and then define the boundaries using our new, larger sample. \\
\indent Fig. 4 shows parallax versus RV for Li-rich (EW(Li)$>150$\,m\AA) sources in our combined sample where sources from \citet{jeffries14} for population A are marked in red, population B in blue and new sources from our observations in green. The separation between the high probability members of each population is clearly apparent in Figure 4, and population A also occupies a smaller range in parallax. 
Based on the clustering apparent in Fig. 3 for sources with significant EW(Li), and on previous estimates of the distance of $\gamma^2$ Vel \citep[e.g. $336_{-7}^{+8}$ pc;][]{north07} and Vela OB2 \citep[$\sim410$pc;][]{dezeeuw99}, we discard sources outside the parallax range $2.2 < \varpi < 3.45$ mas ($\sim290 - 455$ pc) as being unlikely to belong the Vela OB2 region. In Fig. 4 we define a box in parallax and RV around the population A (red) members, within which we select new sources as population A candidates. The edges are defined by 15.69 $<$ RV $<$ 17.69 kms$^{-1}$, which are the values 3$\sigma$ from the median RV of sources in population A (red), and 2.6 $< \varpi <$ 3.2 mas. We also require that for a source to be a member of population A it must lie within the circle in proper motion space illustrated in Fig. 5, of radius 0.6 mas yr$^{-1}$ centered on ($\mu_{\alpha}, \mu_{\delta}$) = (-6.532, 9.753) kms$^{-1}$. This selection circle was chosen as it includes the majority of population A members identified by \citet{jeffries14}. Any other Li-rich target that is not located in both the parallax \& RV box and proper motion circle is assigned to population B. \\
\indent After this selection process we find 57 (26.4\%) sources consistent with being members of population A and 159 members of population B, in contrast to the results of \citet{jeffries14} who allocate 73 (52.5\%) sources to population A and 66 to population B. We end up with fewer sources in population A than \citet{jeffries14} due to imposing tighter restrictions on the membership of sources in Population A from proper motion, and not all of their original members are included in our final sample due to our Gaia DR2 astrometry cuts. In our final populations 18 of the 159 population B members are GES sources that were allocated to population A by \citet{jeffries14}, 40 of our population A members are \citet{jeffries14} population A members, 53 of our population B members are \citet{jeffries14} population B members. The other 105 sources are new additions, 17 for population A and 88 for population B. \\
\indent Fig. 5 shows histograms of the proper motions, RV and parallaxes of our final sample sources, with 3$\sigma$ outliers from the sample median removed, with population A members in red and population B members in blue. The clustered population A stands out as the peak at RV $\approx$ 17 masyr$^{-1}$ and $\mu_{\delta}\approx$ 9.8 mas yr$^{-1}$, but the distinction is not clear in $\mu_{\alpha}$ or parallax where the two populations largely overlap. \\ 
\indent Seven sources allocated to population A that lie outside the GES field are apparent in Fig. 6, though, due to the overlap of the two populations seen in Fig. 5, these may in fact be population B members. Otherwise, the majority of population A members are located within the original GES field, confirming the suggestion made by \citet{jeffries14} that this is a much more compact population than the widely spread population B. \\

\begin{figure}
\includegraphics[width=\columnwidth]{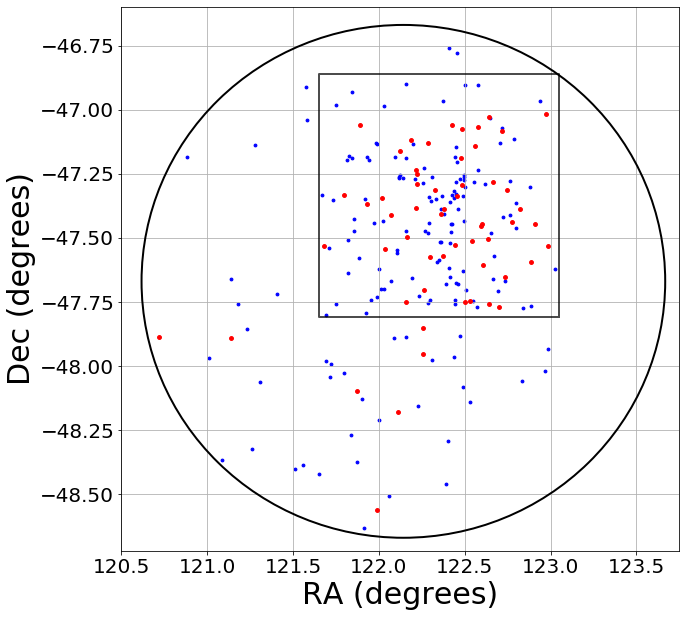}
\setlength{\belowcaptionskip}{-10pt}
\setlength{\textfloatsep}{0pt}
\caption{Positions of 57 population A (red) and 159 population B (blue) members from our final sample. The majority of population A members lie within the 0.9 degree square area observed by GES \protect\citep{jeffries14} but we also identify 7 new population A members further south. }
\label{CSP}
\end{figure}

\subsection{Expansion trends}
Using Gaia DR2 positions, parallaxes, proper motions and our combined sample of RVs, we estimate positions X,Y,Z and velocities U,V,W in the Galactic cartesian coordinate system using a Bayesian inference method. This is done by forward modeling the observed equatorial coordinates, parallaxes, proper motions and RVs from the modeled positions and velocities and the coordinate transformation matrices from \citet{johnson87}. To sample the posterior distribution function we use the Markov Chain Monte Carlo (MCMC) sampler \textit{emcee} \citep{emcee}. For each star we perform 1000 iterations with 100 walkers in an unconstrained parameter space with flat and wide priors. We discard the first half of our iterations as a burn in and from the second half we report the medians of the posterior distribution function as the best fit and use the 16th and 84th percentiles as the 1 sigma uncertainties. See \citet{wright18} for more details on this method. \\
\indent This method is preferable to calculating X,Y,Z,U,V,W from the measured quantites since measurement uncertainties are correlated and distance uncertainties, if derived from parallaxes, are not distributed as Gaussian \citep{bailerjones15}. \\ 
\indent In Figure 7 we show positions X,Y,Z against velocities U,V,W for members of populations A and B in our sample. We calculate best fitting linear relationships between these quantities using MCMC to fit linear relationships between position and velocity. The gradients and their uncertainties for each combination of position against velocity are given in Table. 1.\\
\indent For X versus U, Y versus V and Z versus W, positive or negative gradients are an indication of expansion or contraction of the group \citep{blaauw64}. We find evidence of expansion for population B of at least 4$\sigma$ significance in all three directions (gradients of $ 0.098_{-0.022}^{+0.021}$, $ 0.044_{-0.007}^{+0.007}$, $ 0.069_{-0.011}^{+0.011}$ kms$^{-1}$/pc), but this expansion is significantly anisotropic, the rate of expansion in the X direction being more than twice the rate in the Y direction. Using a two-tailed $z$ test we establish that the difference between the largest and smallest of these gradients is of at least $5\sigma$ significance. We also find some evidence of expansion for population A in the X and Z directions ($ 0.091_{-0.044}^{+0.046}$, $ 0.026_{-0.023}^{+0.022}$ kms$^{-1}$/pc). \\
\begin{figure*}
\begin{center}
    \includegraphics[width=450pt]{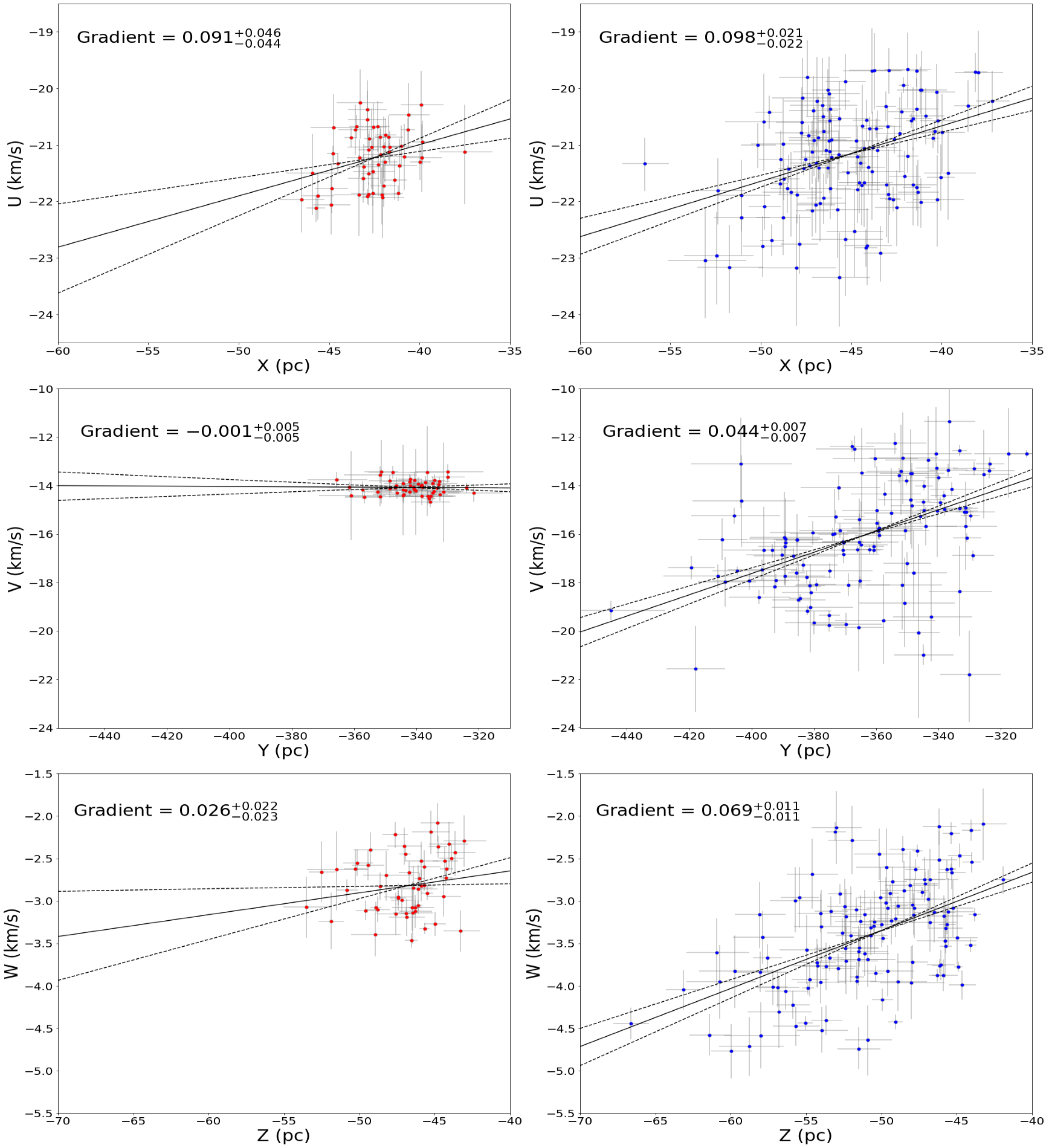}
    \setlength{\belowcaptionskip}{-10pt}
    \setlength{\textfloatsep}{0pt}
    \caption{Cartesian position - velocity plots of populations A (red) and B (blue) with MCMC best fit correlation gradients and uncertainties plotted as solid and dashed lines centered on the mean values of each axis. Note that the ranges plotted in each row are different due to the different dispersions along each axis, but we kept the same range for plots along the same axes so the gradients can be compared.}%
    \label{SelectionBoxes}%
\end{center}
\end{figure*}
\indent Due to $\gamma$ Vel's position in Galactic longitude ($\sim$263$^{\circ}$) the line-of-sight correlates closely to the Y direction, hence we would expect parallax uncertainty to contribute significantly to the estimation of X, Z, U and W. This could create correlations between X and U values and Z and W values that would appear as signatures of expansion in those directions. We attempt to investigate the effect of this covariance in our Cartesian positions and velocities by generating a sample of 1000 stars with Gaussian X,Y,Z,U,V,W distributions defined by the mean and standard deviations of these values for our population A members. We use the coordinate transformation matrices from \citet{johnson87} to calculate positions, parallaxes, proper motions and RVs for this sample and then add random parallax, proper motion and RV uncertainties from Gaussian distributions with the standard deviations of these uncertainty values for our population A members. We then calculate Cartesian positions and velocities by inverting the previous coordinate transformation matrices. We find in fact that the contributions to the position-velocity correlations from correlated uncertainties are small in comparison to our measured gradients ($<$ 0.01 kms$^{-1}$/pc) and do not change the significance levels of the expansion signatures. \\

\subsection{Cluster rotation} 
Rotation is evidenced by correlations between positions X,Y,Z and velocities U,V,W in different directions. There is some evidence for rotation in population A in several dimensions (see Table. 1) but the most significant signature is found in Y vs U at 3$\sigma$ significance ($ 0.029_{-0.009}^{+0.008}$ kms$^{-1}$/pc,Fig. 8). However, interpreting signatures of rotation is more complex than linear expansion or contraction, the same motion may have signatures in multiple dimensions depending on the orientation of the axis of rotation, so we are hesistant to draw physical conclusions from this. Rotation in bound clusters has been observed previously but not frequently. In \citet{henault-brunet12} evidence for rotation was discovered in the cluster R136, and it was argued that clusters may form with at least $\sim$20$\%$ of their kinetic energy in rotation. It will be difficult to put a precise angular velocity to the $\gamma$ Vel cluster without further data and modelling. \\

\begin{figure*}
\begin{center}
    \includegraphics[width=\columnwidth]{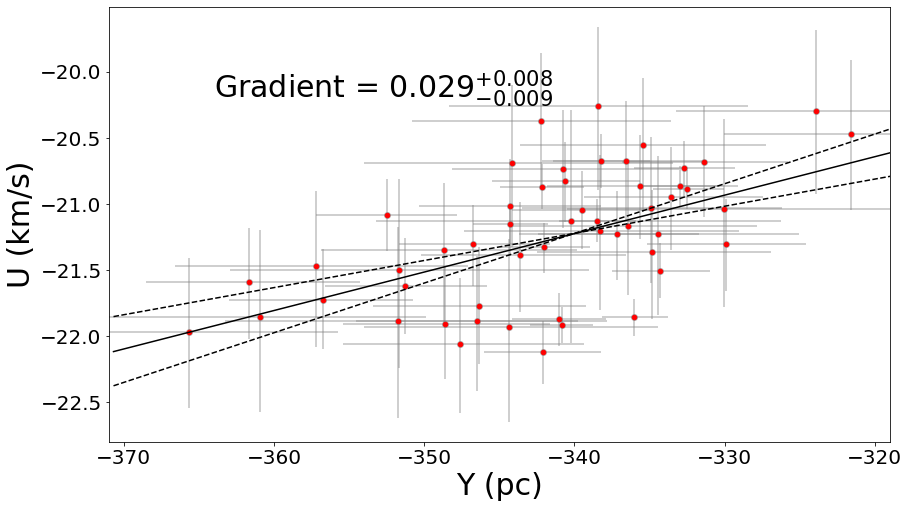}
    \setlength{\belowcaptionskip}{-10pt}
    \setlength{\textfloatsep}{0pt}
    \caption{Cartesian position Y - velocity U of population A with MCMC best fit correlation gradient and uncertainty plotted as solid and dashed lines centered on the mean values of each axis. The significant (3$\sigma$) positive gradient is strong evidence of rotation in this direction.}%
    \label{SelectionBoxes}%
\end{center}
\end{figure*}

\section{Discussion}
The results of the previous section strengthen the hypothesis that population A belongs to the $\gamma$ Vel cluster and that population B belongs to the wider Vela OB2 association, and has interesting implications for the possible formation and evolution mechanisms of these groups.\\
\indent If population B is indeed part of the wider Vela OB2 association, the expansion trends in each dimension would be expected following residual gas expulsion. According to some models \citep[e.g.,][]{baumgardt07}, such expansion trends are expected to be isotropic, but the velocity gradients of this group are in fact strongly anisotropic. However, more recent studies  \citep[e.g.,][]{kruijssen12} suggest that residual gas expulsion may not necessarily produce isotropic expansion patterns and therefore more theoretical work exploring the predicted expansion patterns due to residual gas expulsion are needed. Numerical simulations of residual gas expulsion will be needed to determine whether this mechanism can produce the kinematic behaviour we have found.  \\ 
\indent Such strong evidence for expansion in an association is by no means commonplace. Other recent studies using Gaia astrometry have not found evidence for expansion in other associations \citep[e.g.,][]{wright16,wright18,ward18}. \citet{cantatgaudin19b} also studied the Vela complex and identified signatures of anisotropic expansion in many of the populations present there. Unlike the previously mentioned studies, \citet{cantatgaudin19b} used the unsupervised classification scheme UPMASK to differentiate between multiple populations in their sample differing in position, proper motion and parallax. Likewise, we have used the two-component model of the $\gamma$ Vel population from \citet{jeffries14} to separate two kinematically distinct populations in our sample. The results from these studies may indicate a need to distinguish subgroups present in associations in order to detect the kinematic signatures of expansion that exist. \\
\indent If we instead treat our sample as one group, rather than dividing it into two populations, we still find significant signatures for expansion in each dimension, as we identified for population B. \\

\section{Summary}
We have selected a sample of likely PMS stars in a 2-degree diameter area in the vicinity of the $\gamma$ Vel cluster using Gaia photometry and obtained spectroscopic RVs and EW(Li) measurements for 248 of them. We combine these with the GES $\gamma$ Vel field sample \citep{jeffries14} located within the area of our new observations and with Gaia DR2. We separate the sample into the two populations identified by \citet{jeffries14} using RVs, proper motions and parallaxes. Seven population A members lie outside the GES field, but the majority of population A is located within the smaller GES field, while population B is spread across the whole field. \\ 
\indent We find significant signatures of expansion for population B in all 3 dimensions, which fits with the idea that this population is part of the wider, unbound Vela OB2 association which is in the process of expanding. The rates of expansion in each dimension are also found to be significantly asymmetric. \\
\indent For population A there is no significant signature of expansion in Y or Z directions, which fits with this population belonging to a potentially bound $\gamma$ Vel cluster, though there is a signature of expansion in the X direction. There is some evidence for rotation, with the most significant signature present in Y vs U. \\ 
\indent In order to determine the likely evolution scenario responsible for the asymmetric expansion we have found in this study, and to identify kinematic signatures in stellar populations across the wider area of the Vela complex, a large scale spectroscopic survey over the area of the Vela OB2 association will be necessary to confirm youth and combine with Gaia to give 6D kinematics. 

\section{Acknowledgments}
We thank the referee for their time and a helpful referee's report that has improved this paper. N.J.W acknowledges an STFC Ernest Rutherford Fellowship (grant number ST/M005569/1). This work has made use of data from the ESA space mission Gaia, processed by the Gaia Data Processing and Analysis Consortium (DPAC). Funding for DPAC has been provided by national institutions, in particular the institutions participating in the Gaia Multilateral Agreement. This research made use of the Simbad and Vizier catalogue access tools (provided by CDS, Strasbourg, France), Astropy \citep{astr13} and TOPCAT \citep{tayl05}.

\begin{table}
\begin{center}
\begin{tabular}{|p{0.9cm}|p{0.9cm}|p{2.1cm}|p{2.1cm}| }
\hline
Velocity & Position & Pop A Gradient (kms$^{-1}$/pc) & Pop B Gradient (kms$^{-1}$/pc) \\
\hline
U & X & $ 0.091_{-0.044}^{+0.046}$ & $ 0.098_{-0.022}^{+0.021}$ \\
V & X & $ -0.027_{-0.026}^{+0.025}$ & $ 0.298_{-0.052}^{+0.051}$ \\
W & X & $ 0.037_{-0.030}^{+0.029}$ & $ 0.084_{-0.015}^{+0.015}$ \\
U & Y & $ 0.029_{-0.009}^{+0.008}$ & $ 0.009_{-0.003}^{+0.003}$ \\
V & Y & $ -0.001_{-0.005}^{+0.005}$ & $ 0.044_{-0.007}^{+0.007}$ \\
W & Y & $ 0.002_{-0.005}^{+0.006}$ & $ 0.014_{-0.002}^{+0.002}$ \\
U & Z & $ 0.030_{-0.033}^{+0.033}$ & $ 0.033_{-0.016}^{+0.015}$ \\
V & Z & $ 0.003_{-0.022}^{+0.023}$ & $ 0.230_{-0.037}^{+0.037}$ \\
W & Z & $ 0.026_{-0.023}^{+0.023}$ & $ 0.069_{-0.011}^{+0.011}$ \\
\hline
\end{tabular}
\end{center}
\setlength{\belowcaptionskip}{-10pt}
\setlength{\textfloatsep}{0pt}
\caption{Gradients of MCMC linear best fit models for both A and B populations of the $\gamma$ Vel cluster for every combination of cartesian position and velocity dimensions, as well as uncertainties given by the 16\% and 84\% percentiles of MCMC fits.  }
\end{table}

\bibliographystyle{mn2e}
\bibliography{gammaVelDynamics}
\bsp

\end{document}